\documentclass[aps,twocolumn,groupedaddress,nofootinbib,amsmath,amssymb]{revtex4}
\usepackage{dsfont}
\usepackage{bm}
\begin{document}

\title{Microscopic derivation of  the Schwarzschild  black hole entropy}
\author{Mokhtar Hassaine\thanks{
emails: hassaine-at-inst-mat.utalca.cl} $\medskip \medskip $ \\
{\small \emph{Instituto de Matem\'atica y F\'{\i}sica, Universidad
de Talca, Casilla 747, Talca, Chile.}} }

\begin{abstract}
The main part of this work is to present a formula allowing a
microscopic derivation of the Schwarzschild black hole entropy in
arbitrary dimension. More generally, this Cardy-like formula applies
for static black holes whose gravitational entropy scales as a power
$\alpha$ of the temperature, and is also effective for negative heat
capacity solutions $\alpha<0$. The formula involves the scaling
power $\alpha$, the black hole mass and the energy of a
gravitational soliton identified as the ground state of the theory.
The robustness of this formula is verified in the most famous
example of solution with negative heat capacity, namely the
Schwarzschild black hole. The mass of the Schwarzschild  regular
soliton is computed using the counterterm method for asymptotically
flat spacetimes. Corrections of the black hole entropy of the order
of logarithm of the area are shown to arise for dimensions strictly
greater than four. Finally, we will see that a slight modification
of the Cardy-like formula involving the angular generator, perfectly
reproduces the four-dimensional Kerr entropy.
\end{abstract}

\maketitle

%%%%%%%%%%%%%%%%%%%%%%%
\section{Introduction}
%%%%%%%%%%%%%%%%%%%%%%%%

It is now accepted that black holes are thermodynamical objects
endowed with a temperature and entropy \cite{Bekenstein:1973ur,
Hawking:1974sw}. A natural question has then emerged concerning the
statistical interpretation of the black hole entropy. In the
semi-classical approximation, the entropy is given by the famous
Bekenstein-Hawking formula
\begin{eqnarray}
{\cal S}_{BH}=\frac{c^3 k_B{\cal A}}{4 G\hbar}, \label{BHformula}
\end{eqnarray}
where ${\cal A}$ is the area of the event horizon of the black hole,
$G$ is the Newton's constant, $\hbar$ the Planck's constant, $c$ the
speed of light and $k_B$ the Boltzmann's constant. In what follows,
for simplicity, we will set $c=k_B=\hbar=1$. The presence of the
constants $G$ and $\hbar$ in this formula strongly suggests a
connection between the quantum mechanics and gravity. Nevertheless,
despite the fact that the area dependence of the black hole entropy
(\ref{BHformula}) can be obtained from the quantum partition
function for gravity \cite{Gibbons:1976ue}, its relation to
statistical mechanic is still rather mysterious. In fact, a deep
understanding of the microscopic states responsible of the entropy
area law will certainly provide important insights into quantum
gravity. For this purpose, several interesting ideas have been put
forward in the last decades to provide some possible explanations
concerning the Bekenstein-Hawking entropy. The first convincing
counting of black hole microstates was given by Strominger and Vafa
\cite{Strominger:1996sh} in the case of extremal black holes in
string theory. Later on, this computation was successfully extended
to a wide variety of charged black holes in the extremal and
near-extremal limit, see e. g. \cite{Callan:1996dv,
Horowitz:1996fn}, as well as for non-extremal black holes through
duality relations. There also exist many other microscopic
derivations of black hole thermodynamics such as those exploiting
the AdS/CFT correspondence \cite{Aharony:1999ti} or those involving
the loop quantum gravity \cite{Domagala:2004jt}, the  entanglement
entropy \cite{Bombelli:1986rw} or the asymptotic symmetries
\cite{Strominger:1997eq}; for an exhaustive list, see e. g.
\cite{Carlip:2008wv}. Despite that none of these proposals is
completely satisfactory, these different approaches yield the same
result, raising a problem of "universality" \cite{Carlip:2007ph}. A
certain universality is nevertheless effective in the case of
two-dimensional conformal field theory where the entropy, in the
high-energy regime, is fixed by symmetry independently of any
details, and is given by the famous Cardy formula
\cite{Cardy:1986ie}
\begin{eqnarray}
{\cal S}\sim \ln \rho(\Delta,\bar{\Delta})\sim
2\pi\sqrt{\frac{c_{\tiny{\mbox{eff}}}\Delta}{6}}+
2\pi\sqrt{\frac{\bar{c}_{\tiny{\mbox{eff}}}\bar{\Delta}}{6}}.\label{Cardy}
\end{eqnarray}
Here, $c_{\tiny{\mbox{eff}}}=c-24\Delta_0$ and
$\bar{c}_{\tiny{\mbox{eff}}}=\bar{c}-24\bar{\Delta}_0$ with
$\Delta_0$ and $\bar{\Delta}_0$ being the lowest eigenvalues of the
zero-mode Virasoro generators with central charges $c$ and
$\bar{c}$. As proved by Brown and Henneaux \cite{Brown:1986nw}, much
before the advent of the AdS/CFT correspondence, the asymptotic
symmetries of the three-dimensional AdS space consist in two copies
of the Virasoro algebra with a central charge. Starting from this
observation, it was shown by Strominger \cite{Strominger:1997eq}
that the Cardy formula (\ref{Cardy}) applied for the
three-dimensional BTZ black hole \cite{Banados:1992wn} correctly
reproduces the expression of the Bekenstein-Hawking entropy. This
computation was also generalized in higher dimensions for black
holes having a two-dimensional CFT dual in the case of standard
General Relativity \cite{Guica:2008mu}, and also in presence of
higher-derivative corrections \cite{Azeyanagi:2009wf}.

The deepness of the Cardy formula is essentially due to the modular
invariance which is a characteristic feature of two-dimensional CFTs
only. In order to circumvent this dimensional restriction, a
higher-dimensional extension of Eq. (\ref{Cardy}) was proposed by
Verlinde \cite{Verlinde:2000wg}. This so-called Cardy-Verlinde
formula, which relates the entropy of a CFT with a large central
charge to its total energy and Casimir energy defined as the
sub-extensive part of the energy, has been shown to hold for
strongly coupled field theories having an AdS dual. From an
holographic point of view, the Cardy-Verlinde formula was tested in
various examples admitting AdS vacua with a relative success
\cite{Cai:2001jc}. For completeness, we also mention that a
generalization of the Cardy-Verlinde formula for asymptotically flat
black holes was proposed in Ref. \cite{Klemm:2001pn}.

In the last decade, the interest in extending holography to
spacetimes that are not asymptotically AdS has considerably grown
up. In this context, a Lifshitz version of the Cardy formula was
proposed in \cite{Gonzalez:2011nz} where the $S-$modular invariance
or equivalently the low-high temperature duality arises as a direct
consequence of the isomorphism between the two-dimensional Lifshitz
algebras with dynamical exponents $z$ and $z^{-1}$. Using this
duality, a precise formula for the asymptotic growth of the number
of states was obtained in \cite{Gonzalez:2011nz}, and the robustness
of this formula was tested in a wide variety of Lifshitz black holes
\cite{Gonzalez:2011nz, Ayon-Beato:2015jga, Ayon-Beato:2019kmz}.

More recently, great progress in the microscopic derivation of black
hole entropy has been made possible thanks  to the study of
three-dimensional black holes by highlighting the prominent role
played by the gravitational soliton in the generalizations of the
Cardy formula \cite{Gonzalez:2011nz, Ayon-Beato:2015jga,
Ayon-Beato:2019kmz, Correa:2010hf, Correa:2011dt, Correa:2012rc,
Shaghoulian:2015dwa, Shaghoulian:2015kta, Shaghoulian:2015lcn,
Bravo-Gaete:2015wua, BravoGaete:2017dso}. This point of view is not
in contradiction with the standard Cardy formula where the ground
state is implicitly assumed to be the three-dimensional AdS
spacetime. However, it is important to emphasize that this
assumption is valid only in the vacuum sector and not in the hairy
sector where the ground state may be different. In fact, as shown in
Refs. \cite{Correa:2010hf, Correa:2011dt, Correa:2012rc}, a naive
application of the standard Cardy formula (\ref{Cardy}) for
three-dimensional hairy black holes whose asymptotic symmetry group
remains the conformal one, does not fit with the semi-classical
hairy black hole entropy. This puzzle can be circumvented by first
re-writing the Cardy formula (\ref{Cardy})  in terms of the vacuum
energy rather than the central charge, and by identifying the vacuum
energy with the mass of a gravitational soliton obtained from the
black hole by means of a double Wick rotation as it occurs  for the
AdS soliton \cite{Horowitz:1998ha}. The importance played by the
soliton to correctly reproduce the semi-classical entropy has also
been put in light in the case of anisotropic black brane solutions
\cite{Gonzalez:2011nz, Ayon-Beato:2015jga, Ayon-Beato:2019kmz,
Shaghoulian:2015dwa,Bravo-Gaete:2015wua}, and for AdS black branes
in arbitrary dimension \cite{Correa:2010hf, Correa:2011dt,
Correa:2012rc, Shaghoulian:2015kta, Shaghoulian:2015lcn,
BravoGaete:2017dso}. For these previously mentioned cases, a precise
formula for the density of states at fixed energy ${\cal M}$ was
obtained under the assumption that the bulk gravitational soliton
(identified with the ground state of the putative field theory) has
a negative mass
 denoted by ${\cal M}_{{\tiny{\mbox{soliton}}}}<0$, ensuring in turn the existence of a gap in
the spectrum. We have noticed that the different Cardy-like formulas
reported in Refs. \cite{Gonzalez:2011nz, Ayon-Beato:2015jga,
Ayon-Beato:2019kmz, Correa:2010hf, Correa:2011dt, Correa:2012rc,
Shaghoulian:2015dwa, Shaghoulian:2015kta, Shaghoulian:2015lcn,
Bravo-Gaete:2015wua, BravoGaete:2017dso} in the case of static black
holes\footnote{For rotating AdS black branes, the formula is much
more involved because of the presence of all the rotation
generators, see \cite{Shaghoulian:2015kta,
Shaghoulian:2015lcn,BravoGaete:2017dso}.} can be generically written
as
\begin{eqnarray}
{\cal S}=
\frac{2\pi\,(\alpha+1)}{\alpha^{\frac{\alpha}{\alpha+1}}}\,(-{\cal
M}_{{\tiny{\mbox{soliton}}}})^{\frac{1}{\alpha+1}}\, {\cal
M}^{\frac{\alpha}{\alpha+1}}, \label{genrCardy}
\end{eqnarray}
where the parameter $\alpha>0$ is nothing but the scaling dependence
of the gravitational entropy ${\cal S}_{G}$ with respect to the
temperature, i. e.
\begin{eqnarray}
{\cal S}_{G}\sim T^{\alpha}. \label{scaling}
\end{eqnarray}
This is not surprising since it is known that the entropy of static
anisotropic black branes or of static AdS black branes satisfies
this kind of law temperature dependence (\ref{scaling}) with a
positive scaling $\alpha>0$. This is a direct consequence of the
fact that the horizon metric of these  black holes is an Euclidean
space with zero curvature. Note that for black holes satisfying
(\ref{scaling}), the parameter $\alpha$ is nothing but the quotient
of the heat capacity by the entropy. Hence, the positivity of the
parameter $\alpha$ for solutions satisfying (\ref{scaling}) ensures
as well the positivity of the heat capacity.

For later convenience,  we review the derivation of the generic
formula (\ref{genrCardy}) following closely the steps described in
Ref. \cite{Gonzalez:2011nz}. At low temperature $T={\beta}^{-1}\ll
\epsilon$, the partition function ${\cal Z}[\beta]$ is dominated by
the contribution of the soliton with negative mass ${\cal
Z}[\beta]\approx e^{-\beta {\cal M}_{{\tiny{\mbox{soliton}}}}}$
while at high temperature it becomes
$$
{\cal Z}[\beta]\approx e^{-(2\pi)^{1+\alpha}\beta ^{-\alpha}{\cal
M}_{{\tiny{\mbox{soliton}}}}},
$$
by assuming that the partition function is invariant under the
following generalized modular transformation (with $\alpha>0$)
\begin{eqnarray}
{\cal Z}[\beta]={\cal Z}[(2\pi)^{1+\alpha}\, \beta^{-\alpha}].
\label{modinv}
\end{eqnarray}
In doing so, the asymptotic growth of number of states $\rho({\cal
M})$ at fixed mass ${\cal M}$ given by
\begin{subequations}
\begin{eqnarray}
\label{density}
&&\rho({\cal M})\approx \int e^{f(\beta, {\cal
M})}\,d\beta,\\
&&f(\beta, {\cal M})=\beta{\cal M}-(2\pi)^{1+\alpha}\beta
^{-\alpha}{\cal M}_{{\tiny{\mbox{soliton}}}}, \label{function}
\end{eqnarray}
\end{subequations}
can be evaluated in the saddle point approximation at the extremum
of $f(\beta, {\cal M})$. In this case, the extremum is a minimum
reached at the point
\begin{eqnarray}
\beta_{\ast}=2\pi\, \alpha^{\frac{1}{\alpha+1}}\,\frac{\left(-{\cal
M}_{{\tiny{\mbox{soliton}}}}\right)^{\frac{1}{\alpha+1}}}{{\cal
M}^{\frac{1}{\alpha+1}}}, \label{minimum}
\end{eqnarray}
and, hence the entropy ${\cal S}=\log \rho({\cal M})$ at first
approximation, ${\cal S}\approx f(\beta_{\ast}, {\cal M})$, yields
the formula (\ref{genrCardy}).

It is clear from the expression (\ref{genrCardy}) that the
Cardy-like formula  can be ill-defined for certain negative values
of the parameter $\alpha$. This is not surprising since as shown
previously the derivation of (\ref{genrCardy}), using the modular
transformation (\ref{modinv}),  requires the parameter $\alpha$ to
be strictly positive. Nevertheless, there also exist physical
interesting solutions with negative heat capacity. The most famous
such example is given by the $d-$dimensional Schwarzschild black
hole where the scaling parameter (\ref{scaling}) is given by
$\alpha=2-d<0$. Starting from this observation, we consider as
important the necessity of extending the Cardy-like formula
(\ref{genrCardy}) for negative scaling parameter $\alpha$. We now
see that a similar formula can be obtained for negative values of
$\alpha$ assuming that the mass of the soliton is positive and such
that  $0<{\cal M}_{{\tiny{\mbox{soliton}}}}<{\cal M}$. In this
situation, following the same steps as previously, we end up with
the extremization of the same function (\ref{function}) which is now
located at
\begin{eqnarray}
\beta_{\ast}=2\pi\,\left(-
\alpha\right)^{\frac{1}{\alpha+1}}\,\left(\frac{{\cal
M}_{{\tiny{\mbox{soliton}}}}}{{\cal M}}\right)^{\frac{1}{\alpha+1}}.
\label{maximum}
\end{eqnarray}
As before, at the first approximation  ${\cal S}\approx
f(\beta_{\ast}, {\cal M})$, one will get the following Cardy-like
formula
\begin{eqnarray}
{\cal S}=
\frac{2\pi\,(\alpha+1)}{\alpha}\vert\alpha\vert^{\frac{1}{\alpha+1}}\,({\cal
M}_{{\tiny{\mbox{soliton}}}})^{\frac{1}{\alpha+1}}\, {\cal
M}^{\frac{\alpha}{\alpha+1}}. \label{genrCardyalpha}
\end{eqnarray}
There is however a notable difference with the positive $\alpha$
case. Indeed, for $\alpha<-1$ (as it occurs for the Schwarzschild
solution), the extremum $\beta_{\ast}$ defined by Eq.
(\ref{maximum}) will correspond to a maximum, and consequently
logarithmic corrections to the Bekenstein-Hawking entropy
(\ref{BHformula}) will appear at the next order. More details
concerning this point will be discussed in the end of the paper. For
completeness, we mention that for the remaining negative values of
$\alpha$, namely for $\alpha\in ]-1,0[$, the extremum is a minimum,
and the expression (\ref{genrCardyalpha}) will yield negative values
of the entropy. We will also comment this particular situation in
the last section.

We will test the robustness of the Cardy-like formula
(\ref{genrCardyalpha}) in the case of the $d-$dimensional
Schwarzschild black hole (which is characterized by a negative value
of $\alpha$) by showing that it perfectly reproduces the
Bekenstein-Hawking formula (\ref{BHformula}), that is ${\cal
S}={\cal S}_{BH}$. In order to achieve this task, we will explicitly
construct the Schwarzschild gravitational soliton obtained through a
double Wick rotation followed by a suitable rescaling in order to
avoid conical singularity. In doing so, the resulting configuration
will be shown to be smooth and devoid of any constant of integration
fulfilling what the ground state is expected to be.

The plan of the paper is organized as follows. In the next section,
we will present the $d-$dimensional gravitational soliton and
compute its mass by means of the counterterm method. We will
explicitly check that the Cardy-like formula (\ref{genrCardyalpha})
with a scaling parameter $\alpha=2-d<0$ reproduces the
$d-$dimensional Schwarzschild entropy (\ref{BHformula}).  We will
also show (without any proof) how a slight modification of the
Cardy-like formula (\ref{genrCardyalpha}) perfectly fits with the
four-dimensional Kerr black hole entropy. Comments and concluding
remarks will be reported in the last section. In particular we will
compute the logarithmic corrections of the Schwarzschild black hole
entropy which arise from the Taylor expansion at order two of  the
integrand of the functional given by (\ref{density}).

%%%%%%%%%%%%%%%%%%%%%%%%%%%%%%%%%%%%%%%%%%%%%%
\section{Schwarzschild gravitational soliton}
%%%%%%%%%%%%%%%%%%%%%%%%%%%%%%%%%%%%%%%%%%%%%%
The main objective of the present work is to check the validity of
the Cardy-like formula (\ref{genrCardyalpha}) for black holes
satisfying the condition (\ref{scaling}) with negative values of
$\alpha$. As recalled in the introduction, the negativity of the
scaling parameter $\alpha$ will correspond to black holes with
negative heat capacity. The most famous known of such example is
provided by the Schwarzschild black hole which will be our guiding
example.

In arbitrary dimension $d\geq 4$, we consider the following
regularized action {\small \begin{eqnarray} \label{regAction}
I&=&I_{\tiny{\mbox{bulk}}}+I_{\tiny{\mbox{GH}}}+I_{\tiny{\mbox{ct}}}\\
&=&\frac{1}{16\pi G}\int_M d^dx\sqrt{-g}\,R+\frac{1}{8\pi
G}\int_{\partial M} d^{d-1}x\sqrt{-\gamma}\, K\nonumber\\
&&-\frac{(d-2)}{8\pi G}\int_{\partial M} d^{d-1}x\sqrt{-\gamma}
\,\sqrt{\frac{\tilde{R}}{(d-2)(d-3)}},\nonumber
\end{eqnarray}}
where $K$ is the trace of the extrinsic curvature denoted by
$K_{ab}$, $\gamma_{ab}$ is the induced metric on the boundary
$\partial M$ with Ricci scalar curvature $\tilde{R}$. The
regularized action is composed by the bulk action together with the
standard Gibbons-Hawking boundary term $I_{\tiny{\mbox{GH}}}$, and
the last term is an appropriate counterterm for asymptotically flat
spacetimes \cite{Mann:1999pc,Kraus:1999di}. The Schwarzschild black
hole solution with coordinates $(t,r, x_1,\cdots, x_{d-2})$ is given
by {\small \begin{eqnarray*}
ds^2=-\left(1-\frac{r_h^{d-3}}{r^{d-3}}\right)dt^2+
\frac{dr^2}{1-\frac{r_h^{d-3}}{r^{d-3}}}+r^2 h_{ij}(x) dx^i dx^j,
\end{eqnarray*}}
where $r_h$ is the location of the horizon, and the unit
$(d-2)-$dimensional spherical metric is parameterized as
$$
h_{ij}(x) dx^i dx^j=\sum_{k=1}^{d-3}\left(\prod_{j=k+1}^{d-2}\sin^2
(x_j)\right)dx_k^2+dx_{d-2}^2.
$$
The Schwarzschild mass ${\cal M}$ and gravitational entropy ${\cal
S}_G$ can be computed from the regularized action (\ref{regAction})
yielding the well-known results
\begin{subequations}
\begin{eqnarray}
&&{\cal M}=\frac{(d-2)\,r_h^{d-3}}{16\pi G}\vert
\Omega_{d-2}\vert\label{mass},\\
&&{\cal S}_G=\frac{r_h^{d-2}}{4G}\vert \Omega_{d-2}\vert,
\label{entropy}
\end{eqnarray}
\end{subequations} with $\vert \Omega_{d-2}\vert=\int d^{d-2}x\sqrt{h}$.
On the other hand, the temperature is inversely proportional to the
horizon location
\begin{eqnarray}
T=\frac{(d-3)}{4\pi r_h},
\end{eqnarray}
and consequently the scaling parameter $\alpha$ as defined by Eq.
(\ref{scaling}) is negative and given by $\alpha=2-d$.

In order to check the validity of the Cardy-like formula
(\ref{genrCardyalpha}) with this scaling parameter, we need to
derive the gravitational soliton and compute its mass. The regular
soliton obtained from the Schwarzschild solution by a double Wick
rotation given by $t\to i x_1$ and $x_1\to i t$ followed by a
suitable rescaling that permits to absorb the constant of
integration reads
\begin{widetext}
\begin{eqnarray}
ds^2=&&-\frac{(d-3)^2}{4}\cosh^{\frac{4}{d-3}}(\rho)\prod_{j=2}^{d-2}\sin^2(x_j)dt^2+\cosh^{\frac{4}{d-3}}(\rho)d\rho^2+\tanh^2(\rho)
dx_1^2\nonumber\\&&+\frac{(d-3)^2}{4}\cosh^{\frac{4}{d-3}}(\rho)\left[\sum_{k=2}^{d-3}\left(\prod_{j=k+1}^{d-2}\sin^2
(x_j)\right)dx_k^2+dx_{d-2}^2\right]. \label{soliton}
\end{eqnarray}
\end{widetext}
One can appreciate the regularity of this spacetime (\ref{soliton})
by noting that its Kretchmann invariant
$$
K=R_{\alpha\beta\sigma\gamma}R^{\alpha\beta\sigma\gamma}=\frac{16(d-1)(d-2)^2}{(d-3)^3\cosh^{\frac{4(d-1)}{(d-3)}}(\rho)},
$$
is regular everywhere for $d>3$. After some tedious but
straightforward computations, the mass of the soliton ${\cal
M}_{{\tiny{\mbox{soliton}}}}$ is computed from the regularized
action (\ref{regAction}), once evaluated in the Euclidean
continuation of the soliton (\ref{soliton}), yielding
\begin{eqnarray}
{\cal M}_{{\tiny{\mbox{soliton}}}}=\frac{1}{8\pi
G}\frac{(d-3)^{d-3}}{2^{d-2}}\vert \Omega_{d-2}\vert. \label{delta0}
\end{eqnarray}
As anticipated, we notice that the soliton mass turns out to be
always positive. Now, it is easy to corroborate that the Cardy-like
formula (\ref{genrCardyalpha}) with a scaling parameter
$\alpha=2-d$, and with the black hole and soliton masses
(\ref{mass}-\ref{delta0}) correctly reproduces the Schwarzschild
black hole entropy (\ref{entropy}) or equivalently the
Bekenstein-Hawking formula (\ref{BHformula}).

A legitimate question to ask concerns the extension of the
Cardy-like formula (\ref{genrCardyalpha})  for spinning black holes,
and more particularly for the Kerr solution. It is clear that our
working hypothesis (\ref{scaling}) will  not hold for rotating
solutions. Nevertheless, as we will show below, a slight
modification of the formula (\ref{genrCardyalpha}) involving the
angular momentum can perfectly fit with the four-dimensional Kerr
entropy. Following the notations of Refs. \cite{Caldarelli:1999xj,
Gibbons:2004ai}, where the gravitational constant is fixed to unit,
$G=1$, and without the cosmological constant, i. e. $l\to \infty$,
the Kerr mass ${\cal M}$, its angular momentum ${\cal J}$ and its
entropy are given by
\begin{eqnarray}
{\cal M}=\frac{r_h^2+a^2}{2r_h},\quad {\cal J}=a{\cal M},\quad {\cal
S}_G=\pi(r_h^2+a^2), \label{kerrthermo}
\end{eqnarray}
where $a$ is the rotation parameter. It is intriguing to notice that
our Cardy-like formula (\ref{genrCardyalpha}) in four dimensions,
(corresponding to $\alpha=-2$) with the introduction of the angular
momentum ${\cal J}$ through the change
\begin{eqnarray}
{\cal M}\to\frac{1}{\sqrt{2}} \sqrt{{\cal M}^2+\sqrt{{\cal
M}^4-{\cal J}^2}}, \label{changeM}
\end{eqnarray}
correctly reproduces the Kerr entropy (\ref{kerrthermo}). The
multiplying factor in the expression (\ref{changeM}) ensures that in
the zero spinning case ${\cal J}=0$, one recovers exactly the mass.
The generalization for higher-dimensional Kerr solution which is
clearly beyond the scope of the present work is far from obvious
essentially because, as the dimension increases, the number of
rotations parameters does as well. As a direct consequence, a
substitution of the form (\ref{changeM}) may become much more
involved.

%%%%%%%%%%%%%%%%%%%%%%%%%%%%%%%%%%%%%%%%%%%
\section{Comments and concluding remarks}
%%%%%%%%%%%%%%%%%%%%%%%%%%%%%%%%%%%%%%%%%%
The two formulas (\ref{genrCardy}) and (\ref{genrCardyalpha}) which
apply  respectively for $\alpha>0$ with a negative mass soliton
${\cal M}_{{\tiny{\mbox{soliton}}}}<0$, and for $\alpha<0$ with
${\cal M}_{{\tiny{\mbox{soliton}}}}>0$, can be combined in a single
formula that holds for any non-zero $\alpha$ as
\begin{eqnarray}
{\cal S}=
\frac{2\pi\,(\alpha+1)}{\alpha}\vert\alpha\vert^{\frac{1}{\alpha+1}}\,\vert{\cal
M}_{{\tiny{\mbox{soliton}}}}\vert^{\frac{1}{\alpha+1}}\, {\cal
M}^{\frac{\alpha}{\alpha+1}}. \label{genrCardyalphacompact}
\end{eqnarray}
The robustness of this generic Cardy-like formula
(\ref{genrCardyalphacompact}) for static black holes satisfying the
scaling condition ${\cal S}\propto
T^{\alpha}$ has been verified :\\
(i) previously, for black branes with different asymptotics (AdS,
Lifshitz, hyperscaling violation), and with positive values of the
parameter $\alpha$ in Refs. \cite{Gonzalez:2011nz,
Ayon-Beato:2015jga, Ayon-Beato:2019kmz, Correa:2010hf,
Correa:2011dt, Correa:2012rc, Shaghoulian:2015dwa,
Shaghoulian:2015kta, Shaghoulian:2015lcn,
Bravo-Gaete:2015wua, BravoGaete:2017dso},\\
(ii) and in the present work for negative values of $\alpha$ in the
case of the asymptotically flat Schwarzschild black hole solution.\\
In all these working examples, the gravitational entropy is
proportional to area of the horizon ${\cal S}_G\propto {\cal A}$,
and hence this last condition can be thinking to be necessary for
the validity of the formula (\ref{genrCardyalphacompact}). It will
be interesting to clarify this point by checking the validness or
not of the formula (\ref{genrCardyalphacompact}) for black holes
whose entropy satisfies our working hypothesis (\ref{scaling}), but
without being proportional to the area of the horizon. Of one such
example is provided by a family of asymptotically flat Lovelock
black holes \cite{Crisostomo:2000bb} parameterized by an integer
$k=\left\{1, 2,\cdots, [\frac{d-1}{2}]\right\}$ whose metrics
{\small
\begin{eqnarray*}
ds^2=-&&\left(1-\frac{r_h^{d-2k-1}}{r^{d-2k-1}}\right)dt^2+
\frac{dr^2}{1-\left(\frac{r_h}{r}\right)^{d-2k-1}}\nonumber\\
&&+r^2 \left[\sum_{p=1}^{d-3}\left(\prod_{j=p+1}^{d-2}\sin^2
(x_j)\right)dx_p^2+dx_{d-2}^2\right],
\end{eqnarray*}}
are solutions to the field equations associated to the
$d-$dimensional actions labeled by the integer $k$, and defined by
\begin{eqnarray*}
I_k=\int_{M} \epsilon_{a_1\cdots a_d}R^{a_1a_2}\cdots
R^{a_{2k-1}a_{2k}}e^{a_{2k+1}}\cdots e^{a_d}.\\
\end{eqnarray*}
Here $e^a$ is the vielbein,
$R^{ab}=d\omega^{ab}+\omega^a_{\,c}\omega^{cb}$ stands for the
curvature two-form, and the wedge products between the forms are
voluntary omitted for simplicity. For these asymptotically flat
Lovelock solutions, the gravitational entropy is given by ${\cal
S}\propto r_h^{d-2k}$, and hence it will be proportional to the area
of the horizon only in the Einstein-Hilbert case $k=1$. In a
forthcoming work, we will precisely study the microscopic derivation
of Lovelock black hole entropy in the asymptotically flat case as
well as in presence of a cosmological constant
\cite{workinprogress}. The interest of such work is also motivated
by the fact that the Cardy-Verlinde formula \cite{Verlinde:2000wg}
has been shown to fail for the AdS Lovelock black holes
\cite{Cai:2001jc}, even in the large horizon radius limit where the
AdS Lovelock entropy becomes proportional to the area of the
horizon. Note that in this regime, i. e.  $r_h \gg \epsilon$, the
AdS Lovelock entropy satisfies as well our working hypothesis
(\ref{scaling}), and consequently it will be interesting to check
the validity of the formula reported here
(\ref{genrCardyalphacompact}).

Another point to be clarified is that for $\alpha<0$ in the range
$\alpha\in ]-1,0[$, the expressions (\ref{genrCardyalpha}) or
(\ref{genrCardyalphacompact}) will yield negative values of the
entropy. One could hence think that the range of the possible values
of $\alpha$ must exclude the interval $]-1,0[$, but fortunately
there exist black hole solutions with negative entropy in the
current literature, see e. g. \cite{Hassaine:2015ifa,
Chernicoff:2016jsu, Cisterna:2019uek}. One of such example is given
by a solution of the  five-dimensional vacuum Einstein equations,
$R_{\mu\nu}=0$, with a horizon topology modeled by the so-called Nil
geometry (which is one the eight geometries of the three-dimensional
Thurston classification \cite{Thurston}). More precisely, the Nil
black hole metric, solution of the vacuum Einstein equations,  is
given by \cite{Hassaine:2015ifa} {\small
\begin{eqnarray}
ds^2=-(1-r_h/r)dt^2&&+\frac{dr^2}{r^5(1-r_h/r)}+\frac{1}{r}(dx_1^2+dx_2^2)\nonumber\\
&&+r(dx_3-x_1dx_2)^2, \label{bhthurston}
\end{eqnarray}}
and its mass, entropy and temperature computed by the Euclidean
method yield \cite{Hassaine:2015ifa}
\begin{eqnarray}
{\cal M}= \frac{\vert\Omega_3\vert}{32\pi G}r_h,\quad {\cal
S}=-\frac{\vert\Omega_3\vert}{4G\,\sqrt{r_h}},\quad
T=\frac{r_h^{\frac{3}{2}}}{4\pi}. \label{thurstonsol}
\end{eqnarray}
It is clear from these expressions that  the scaling parameter as
defined by Eq. (\ref{scaling}) is given by $\alpha=-\frac{1}{3}$,
and hence from the multiplying factor of the entropy formula
(\ref{genrCardyalpha}), namely $(\alpha+1)/\alpha$, this would yield
to a negative entropy in complete accordance with
(\ref{thurstonsol}). In addition, one can note that the mass term in
the Cardy-like formula (\ref{genrCardyalpha}), i. e. ${\cal
M}^{\alpha/(\alpha+1)}$, will reproduce the correct horizon
dependence of the entropy ${\cal S} \propto r_h^{-1/2}$. Now, in
order to fully corroborate the formula (\ref{genrCardyalpha}), it
will remain to identify the ground state and to compute its mass.
However, it is evident from the Nil black hole solution
(\ref{bhthurston}) that any Wick rotation will drastically change
the signature of the solution. Consequently, the recognition of the
appropriate ground state which seems to be far from obvious should
deserved some attention.

We have also seen that in four dimension, a slight modification
given by (\ref{changeM}) in the Cardy-like formula
(\ref{genrCardyalpha}) reproduces the Kerr entropy. It will be
interesting to have a general formula that fits with the Kerr
entropy in arbitrary dimension $d$ involving all the different
rotation generators ${\cal J}_i$, with $1\leq i\leq [(d-1)/2]$ .

Finally, another interesting problem is concerned by the subleading
corrections of the black hole entropy (\ref{BHformula}) which are
known to be proportional to the logarithm of the area
\cite{Kaul:2000kf}. For negative values of $\alpha$, we have seen
that $\beta_{\ast}$ defined by Eq. (\ref{maximum}) corresponds to a
maximum of the function $f$ defined in (\ref{function}).
Consequently,  a Taylor expansion at second order of the integrand
of the asymptotic growth of number of states (\ref{density}) will
yield
$$
{\cal S}\approx f(\beta_{\ast},{\cal
M})-\frac{1}{2}\log\left[\frac{\alpha\,(\alpha+1)\,{\cal
M}^{\frac{\alpha+2}{\alpha+1}}}{\left({\cal
M}_{{\tiny{\mbox{soliton}}}}\right)^{\frac{1}{\alpha+1}}}\right],
$$
and, in the Schwarzschild case, this expression becomes
\begin{eqnarray}
{\cal S}\approx \frac{{\cal
A}}{4G}-\frac{1}{2}\left(\frac{d-4}{d-2}\right)\log({\cal
A})-\frac{1}{2}\log(\gamma), \label{logcorrections}
\end{eqnarray}
where $\gamma$ is an area-independent factor given by
$$
\gamma=\frac{(d-2)(d-3)^2}{32\pi
G}\,\vert\Omega_{d-2}\vert^{\frac{2}{d-2}}.
$$
It is interesting to note that the corrections of order of the
logarithm of the area only appear for dimensions strictly greater
than four. Indeed, for the four-dimensional Schwarzschild metric,
the entropy approximation (\ref{logcorrections}) reduces to
$$
{\cal S}\approx \frac{{\cal
A}}{4G}-\frac{1}{2}\log\left(\frac{1}{4G}\right).
$$
\\
\\
\bigskip

\textbf{Acknowledgments.-}  This work is partially supported by
grant from ECOS-CONICYT 180011.

%%%%%%%%%%%%%%%%%%%%%%%%%%%

\end{document}